# PREDICTING THE STUDENTS INVOLVEMENTS AND IT'S IMPACTS ON LEARNING OUTCOMES THROUGH ONLINE EDUCATION DURING COVID-19

**Muhammad Nadeem**
Computer Science Department, University of the Punjab

**Faisal Bukhari**
Data Science Department, University of the Punjab

**Ali Hussain**
Computer Science Department, University of the Punjab

**Abstract**
Everybody knows very well about the COVID-19 pandemic, lockdown, and its impacts and effects on every field of life, from childhood to senior citizens, from local to global. The underlying research study focuses on students' involvement in online classes. This paper assesses the effect of the COVID-19 pandemic on the students' participation and involvement during online classes compared to the physical classes, cheating behavior, health effects, and study styles of the students of diverse degrees and age groups. This research study contributes to the real problems and challenges that students faced during online classes during the COVID-19 pandemic. The percentages of the students' responses with different color schemes shown in Fig. 1, Fig. 2, Fig.3(a), Fig.3(b) and Fig.4 are conveying powerful and meaningful insight. These figures and the results given in Table I and Table II indicate that most students are not fully involved during online classes due to technical issues, remote distance, etc. We applied the Test here because we do not have exact population means. We used ttest_1samp with default value 0 to compute the variables' statistics and p-value. These values are minimal in favor of rejecting the null or H0 (hypothesis) and accepting the alternate or H1 (hypothesis). It further means that students' involvement during online classes is severely affected.

## I. INTRODUCTION

The primary motivation for selecting this topic is that the quality of education is directly proportional to the involvement of the students during the lecture. Firstly, I found it as a teacher that many students have left the online lecture physically, but logically they showed their status as a present. This problem has multiple issues. The respected teacher cannot be confident about the presence of the students physically during online lectures. Secondly, the students are facing different issues during online lectures. The impact of these issues is that they lose interest in learning during online lectures. This research work is a new study focused mainly on the level of student involvement during online lectures. All the countries attacked by the villainous COVID-19 virus that has upset each area of life as per economy, from producers to consumers [1]. During the Covid19 pandemic, the Education sector was also severely impacted. The forceful impact of this virus sent the students and teachers to study and teach remotely from face to face system of education. Resultantly, Educational institutions are searching for another way to teach and evaluate the students [2]. So to keep every student and teacher safe, all the Educational Institutions closed because of the citywide, districtwide, and countrywide lockdowns. In such lockup situations, the students and teachers cannot interact face-to-face [3].

To keep the chain of teaching in COVID-19 virus, the World Bank has been actively trying to give financial assistance to the underdeveloped or more affected countries. The ultimate goal of [4] is to provide basic education rights to every student during this viral disease. As far as online learning is concerned, there is much use of technology. This technology-dependent way of education becomes a barrier for learners who did not train to use technology [5]. Similarly, in Pakistan, in 2021, all the educational institutions have closed as the previous year due to the severity of COVID-19. Pakistani Ministry of Education and Higher Education Commission (HEC) also provides online and distance learning ways to teach the students. [6]. The HEC provided the design for online policy guidance notes and guidelines for the Universities. However, It's a reality that practical work is not being taught during online education. This also demotivated the students, and it made an impact on their involvement in online lectures [7]. In addition to the problems mentioned above and issues of students and teachers, there are also the problems of admin staff [8].

Therefore, the teachers are not satisfied with the student's involvement in online classes compared to physical classes.

In this connection, to find the answers, this study would work on the following research objectives:

- To predict why the students are involved is not as much as physical class.
- To find why the students are not interested in attending the full online lecture.
- To discover the issue faced by the students during online lecture.
- To find the impact of taking lectures in class room with the lecture taking online on the students' learning outcomes.
- To find the family members' realization about their children's online study.

The outcomes of the research would be necessary for the following concerning levels:

- Student
- Teacher
- Parents
- Educational Institution
- Education Ministries

The most crucial stakeholder in the learning process are teachers, and students are aware of the issues and the factors involved as per the student involvement during an online class. The parents would also notice the difference in attitude and aptitude to study in the classroom and at home via online education. The Educational Institution may send reports to the Ministry of Education and HEC based on the outcomes of the student's involvement during an online class. In this way, the Ministries can inform the Government to look after the policies to plan a different mature online education system or to open the educational institution as soon as possible.

## II. LITERATURE REVIEW

The impacts of COVID-19 on health, society, and education are highlighted in [9]. The researchers divided their research into four different groups: general demographics, information about daily online routine, assessment of the learning of online experience and level of satisfaction of the students, and evaluation of health due to change in lifestyle. Cheating during the exam is one of the main problems. The research work done by [10] on cheating shows that an individual's strengths vary according to the achievement settings. Their findings also concluded that the cheating rate was higher in educational settings than in work areas and in work sites than in sports venues. Study 1 further suggests that the strengths of individuals' cheating intentions differ across achievement settings.

Intentions to cheat were higher in educational settings than in work settings and higher in work settings than in sports settings. The outcomes of this research [11] concluded that the online examination during COVID-19 increased the cheating ratio, which is unrelated to achievement goals. The studies provided different guidelines to the teachers for setting the questions and time duration for online exams. The researchers of [12] highlighted the levels of students' stress, depressive symptoms, loneliness, effects of missing social life, and specific worries for their undergraduate studies. They also showed extreme crises of the students on health and research during lockdown due to COVID-19. The authors discussed that they got 212 responses out of 266 from students for the crises suffered. They also recommended different plans for teachers and academic institution administrators to develop online events so that they can prepare newcomers very well. The research efforts of [13] discover the critical problems faced by the students in the present e-learning system. They have also found the factors influencing online learning during COVID-19.

The authors also discussed the impacts of students' willingness to study alone in an e-learning environment. In addition, they interviewed 30 students from six Universities and conducted meetings with 31 e-learning system experts to find the main problems. They also suggested applicable plans for policymakers, developers, designers, and researchers, enabling them to be better acquainted with the critical aspects of the e-learning system during the COVID-19 pandemic. The researchers of [14] have found too much dissatisfaction during the online study on the COVID-19 situation. The outcomes of this research concluded that the students of the dental study were dissatisfied with the online teaching during COVID-19. The results of this research crying that online study is disturbing the student's level of involvement in the study very severely. The efforts of the analyses highlighted different aspects of students during the online study in the COVID-19 pandemic worldwide. They discussed and evaluated severe issues such as technical and economic issues, psychological problems, and students' fears about the future. It badly affects the study taste of the students and their pace in the learning process. They also offered different plans and suggestions for the policymakers and higher authorities to overcome the issues faced by the students and the teachers. The research study by the authors of [15] observed and evaluated the impact of the perception of e-learning crashes. They discovered its impact on psychological upset in the students during the COVID-19 pandemic. They concluded that fear of academic loss had become the main reason for mental upset during the issues of online study in corona disease. They also suggested remedies for the policymakers and educational institutions to manage the student's stress during the online study. The researchers analyzed different types of challenges faced by the students in Pakistani Universities [16]. The main obstacles highlighted are economic, technical, lack of skills, family support, etc. They also recommended that the Govt. take a severe step to overcome the challenges faced by the students. The outcomes of this research work [17] show that the students do not want to study online. The students expressed their problems during the survey that they were not prepared and trained for such a learning shift. They do not have a non-stop electricity facility and well-equipped information technology-based infrastructure at their homes.

## III. PROBLEM STATEMENT

To find the effect of the COVID-19 pandemic on the involvement of the students during online classes as compared to the physical classes, cheating behavior, health effects, and study styles from the students of diverse degrees and age groups.

Hypothesis:

H0 = Student’s involvement during online classes is the same as in physical classes.

H1 = Student’s involvement during online classes is not the same as in physical classes.

**Methodology and Data Collection**

The survey methodology used to accomplish this research. Survey is a method for the collection of the information for the sample of individuals [18]. The findings of the survey analyzed through statistical analysis.

**• OBJECTIVES OF THE SURVEY**

To analyze the levels of the student's involvement and its impacts on learning outcomes during online lectures during COVID-19.

**•TARGET POPULATION**

Graduate, Undergraduate and Intermediate students of the Universities and Colleges

**•DATA TO BE COLLECTED**

A questionnaire developed based on the literature review. Then this questionnaire circulated online as much as possible to find the maximum responses from the target population due to the COVID-19 situation.

**•MEASUREMENT `INSTRUMENT'**

The measurement instrument of the required survey is a questionnaire. The questions of this questionnaire were
closed-ended with a Likert scale. The definition of the Likert scale is given below:
1. SA (Strongly Agreed)
2. A (Agreed)
3. U (Undecided),
4. D (Disagreed)
5. SD (Strongly Disagreed)

This questionnaire would be distributed through Google docs to make it available to the targeted population and to get a maximum number of responses.

**IV. DESIGN OF RESEARCH STUDY**

An online survey performed using Google online forms. However, the questionnaire of this survey consists of the following subsections:

A. Respondents will be requested to answer their following usual demographics:
• Age
• Gender
• Area of residence

B. Getting information routine wise online learning during the shift from face to face study to online study in colleges/Universities in Pakistan. These information consists of the following:
• Average time given for online study in hours per day
• Quality and the problems of the communication medium
• Actual involvement in virtual lecture same as face to face lecture in physical class
• Level of interruption by the family members during online study period
• Attention and focus level from joining to the end of online class.
• Effects of online learning on Cheating behavior and students involvement to

C. Evaluation of the experience of the student's level of involvement in virtual class to find the overall students involvement in online lecture.

D. Evaluation of health during change in learning style from physical class environment provided by the College/University to the virtual class environment provided by your parents at home and the effects of virtual class on your involvement of class.

The pictorial survey responses are given below:

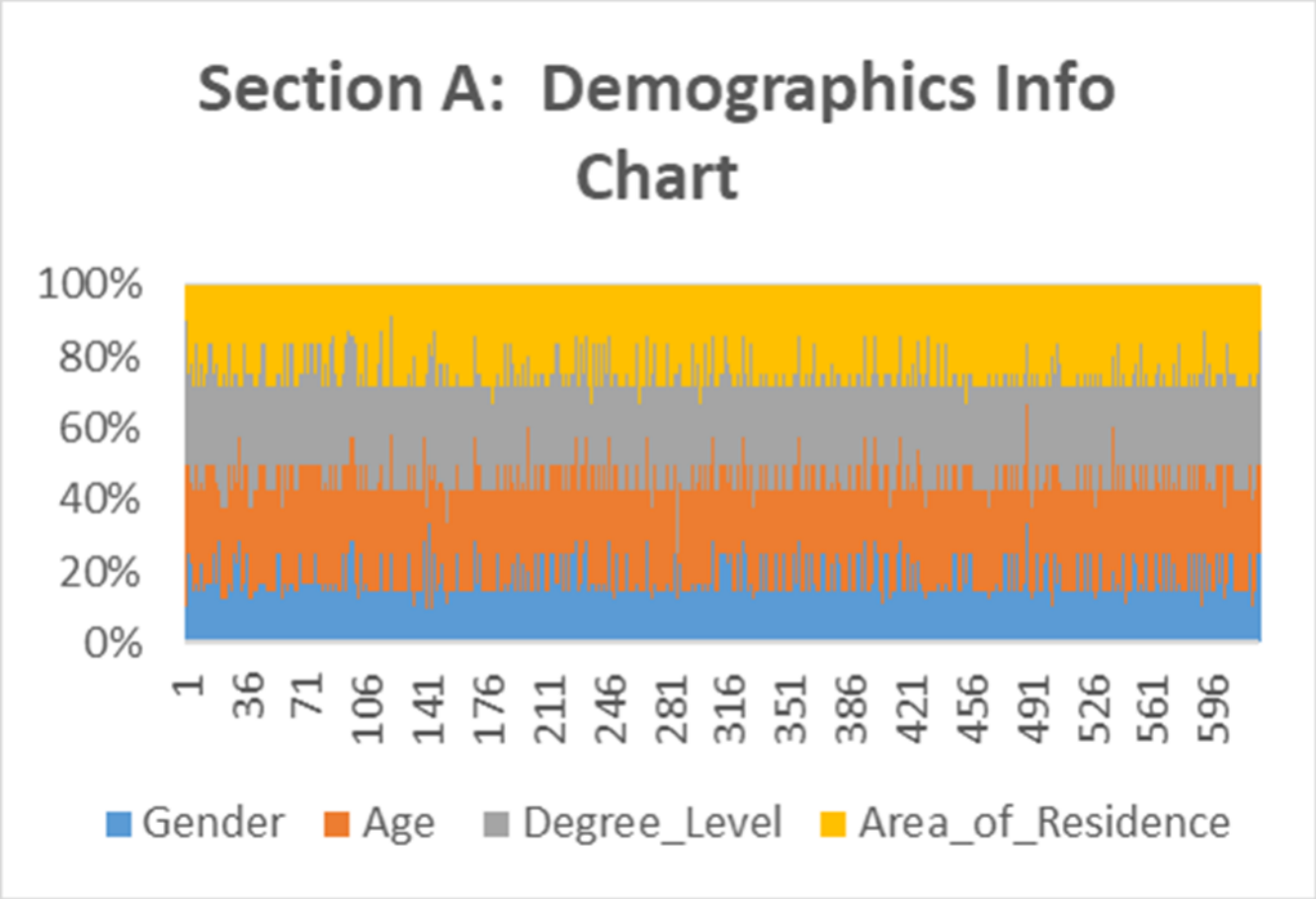

**Fig. 1.** Getting General Info

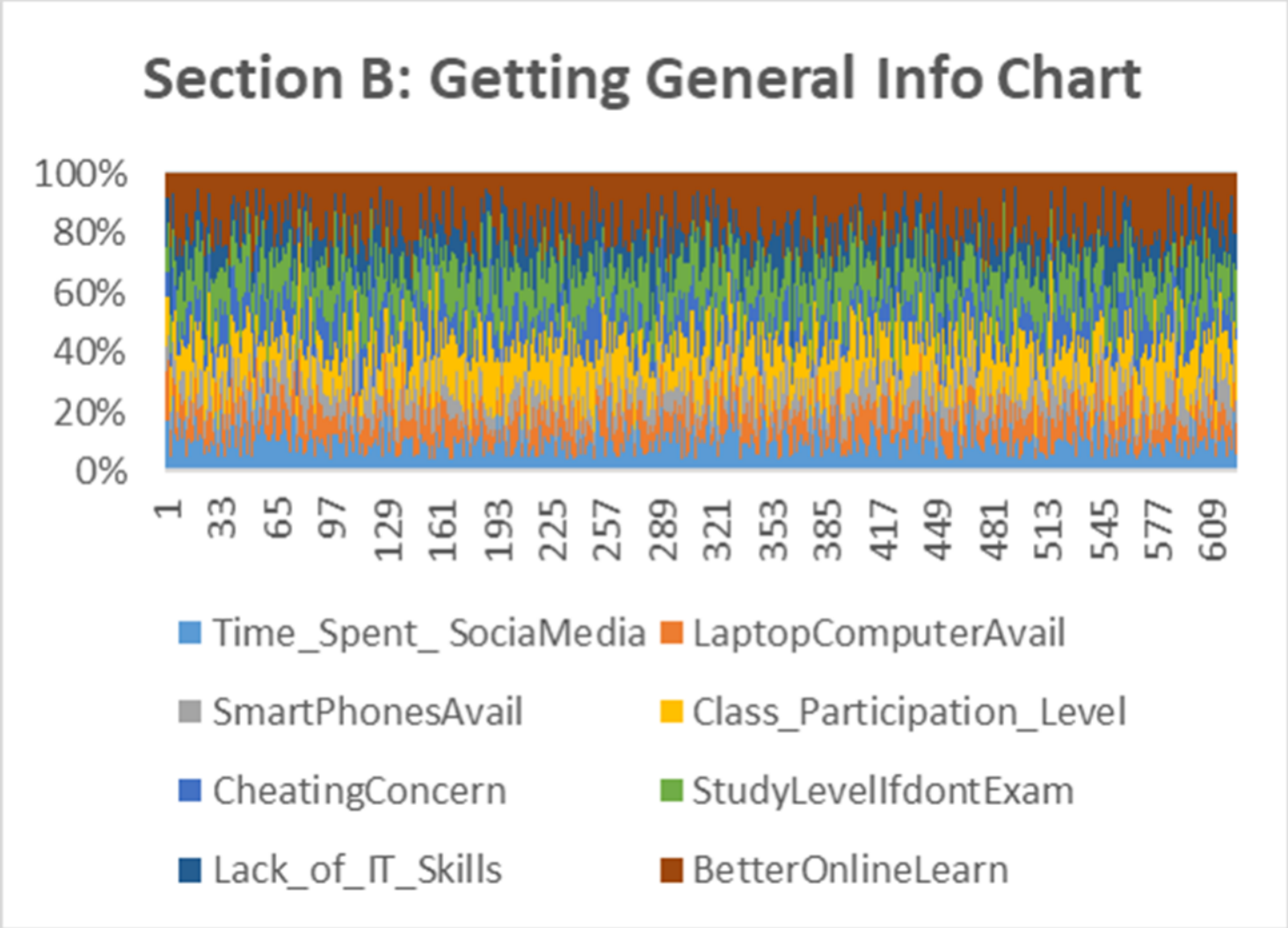

**Fig. 2.** Getting General Info

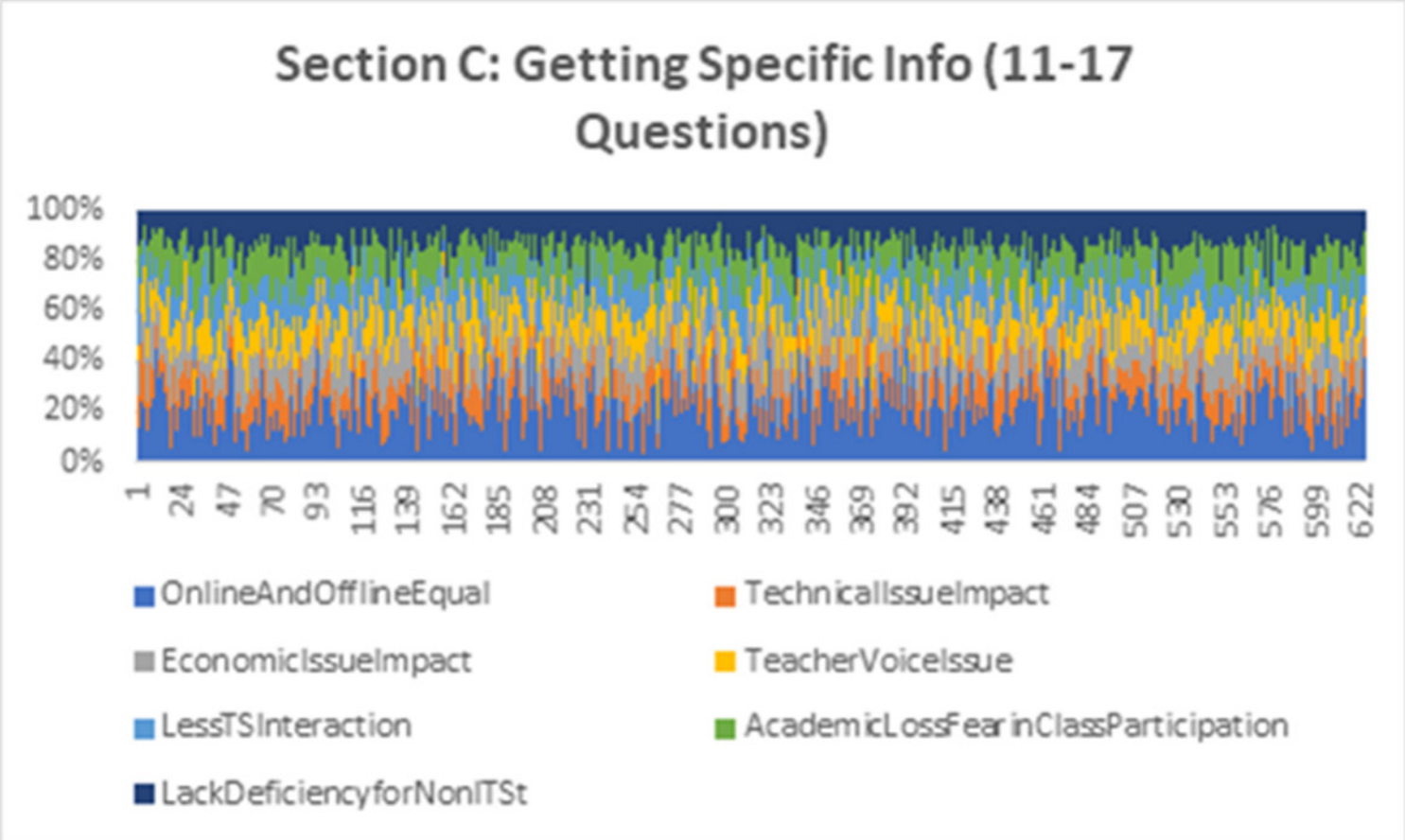

**Fig. 3(a).** Getting Specific Info

We have created questionnaire. Its soft copy is available at the following link: https://docs.google.com/forms/d/1zqnXC9EXRXjmNL7VX2FP4hh6OL0NVu-C_w8QZOFxRsc/edit )
We have collected 623 responses from different level of degree students.

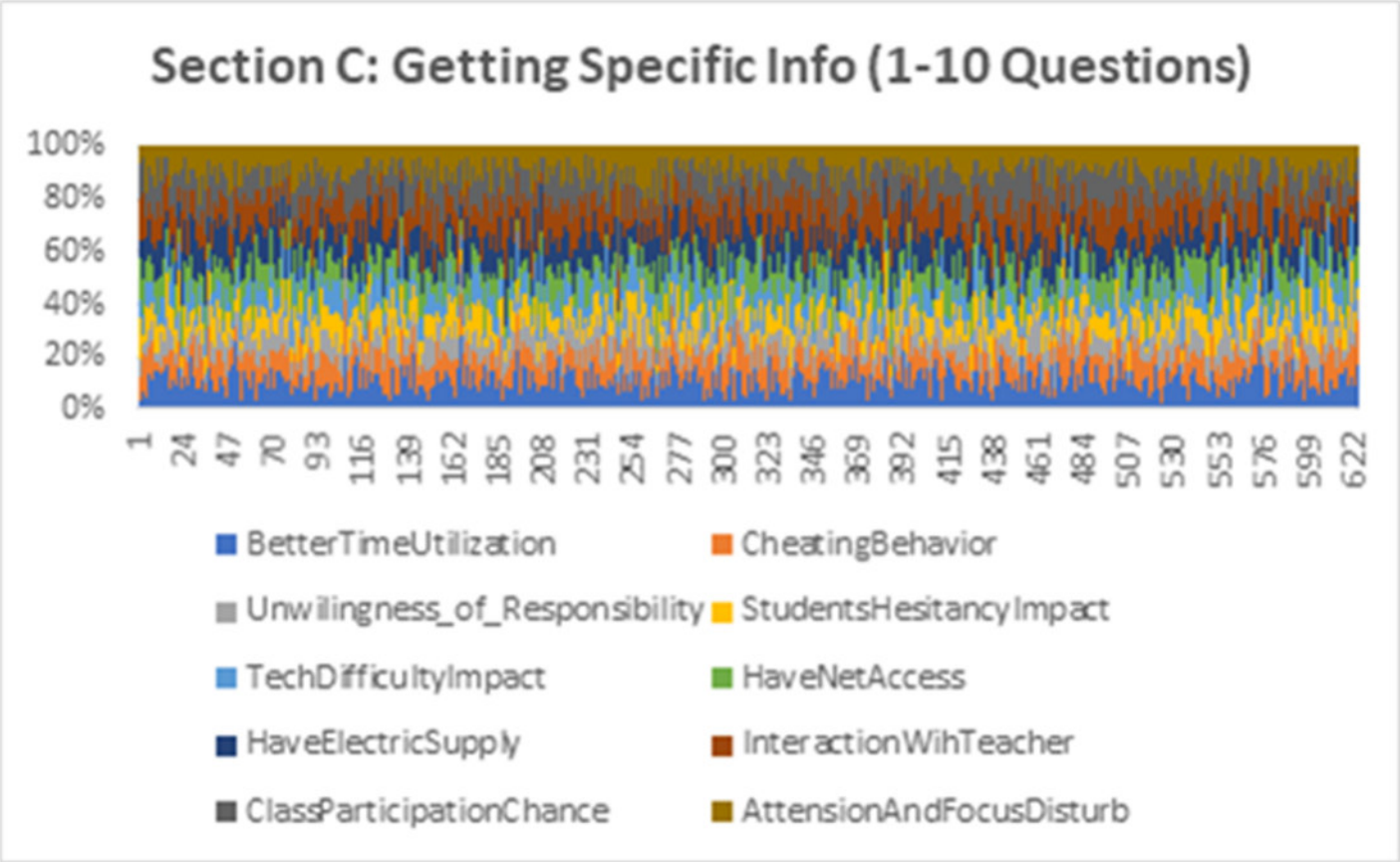

**Fig. 3(b).** Getting Specific Info

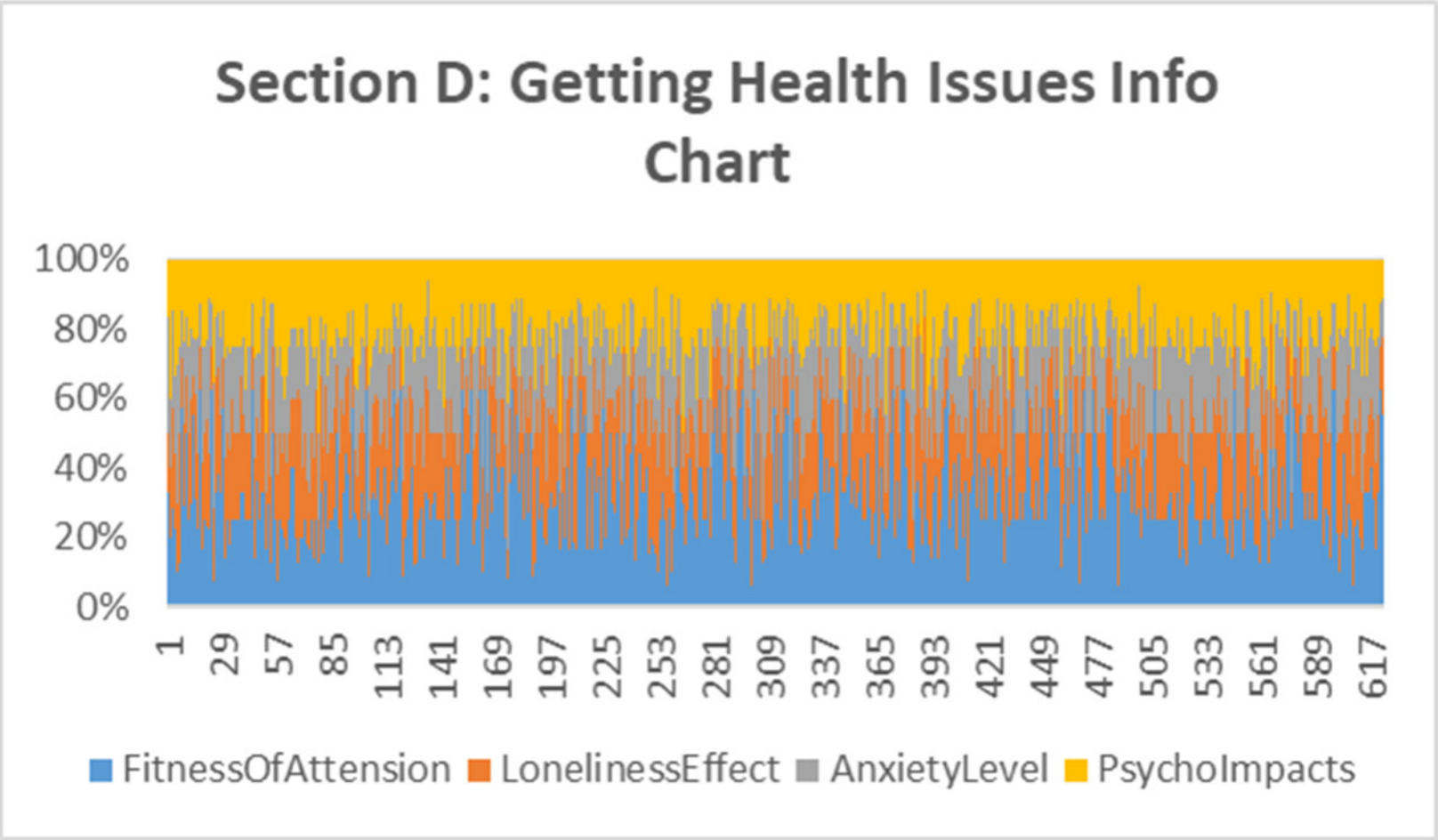

**Fig. 4.** Getting Health Issues info

## V. EXPEIRMENTAL RESULTS

The means and standard deviation of all the variables as per questionnaire are given below:

**TABLE I.**

| S.# | Variable | Value |
|---|---|---|
| **Mean of all the variables** | | |
| SECTION A: DEMOGRAPHICS INFO | | |
| 1 | Gender | 1.400000 |
| 2 | Age | 2.028571 |
| 3 | Degree_Level | 2.257143 |
| 4 | Area_of_Residence | 1.628571 |
| SECTION B: GETTING GENERAL INFO | | |
| 1 | Time_Spent_ SociaMedia | 2.457143 |
| 2 | LaptopComputerAvail | 1.771429 |
| 3 | SmartPhonesAvail | 1.571429 |
| 4 | Class_Participation_Level | 2.885714 |
| 5 | CheatingConcern | 1.942857 |
| 6 | StudyLevelIfdontExam | 3.028571 |
| 7 | Lack_of_IT_Skills | 2.485714 |
| 8 | BetterOnlineLearn | 3.600000 |
| SECTION C: GETTING SPECIFIC INFO | | |
| 1 | BetterTimeUtilization | 3.342857 |
| 2 | CheatingBehavior | 2.285714 |
| 3 | Unwilingness_of_Responsibility | 2.114286 |
| 4 | StudentsHesitancyImpact | 2.371429 |
| 5 | TechDifficultyImpact | 2.200000 |
| 6 | HaveNetAccess | 2.514286 |
| 7 | HaveElectricSupply | 3.000000 |
| 8 | InteractionWihTeacher | 3.085714 |
| 9 | ClassParticipationChance | 3.057143 |
| 10 | AttensionAndFocusDisturb | 2.342857 |
| 11 | OnlineAndOfflineEqual | 3.800000 |
| 12 | TechnicalIssueImpact | 1.857143 |
| 13 | EconomicIssueImpact | 2.000000 |
| 14 | TeacherVoiceIssue | 1.971429 |
| 15 | LessTSInteraction | 1.857143 |
| 16 | AcademicLossFearinClassParticipation | 2.028571 |
| 16 | AcademicLossFearinClassParticipation | 0.970588 |
| 17 | LackDeficiencyforNonITSt | 0.747240 |
| **Standard Deviation of all the variables** | | |
| SECTION A: DEMOGRAPHICS INFO | | |
| 1 | Gender | 0.489898 |
| 2 | Age | 0.376883 |
| 3 | Degree_Level | 0.552545 |
| 4 | Area_of_Residence | 0.483187 |

| SECTION B: GETTING GENERAL INFO | | |
|---|---|---|
| 1 | Time_Spent_ SociaMedia | 1.078169 |
| 2 | LaptopComputerAvail | 0.897161 |
| 3 | SmartPhonesAvail | 0.766652 |
| 4 | Class_Participation_Level | 1.259738 |
| 5 | CheatingConcern | 1.093954 |
| 6 | StudyLevelIfdontExam | 1.502107 |
| 7 | Lack_of_IT_Skills | 1.105090 |
| 8 | BetterOnlineLearn | 1.515633 |
| SECTION C: GETTING SPECIFIC INFO | | |
| 1 | BetterTimeUtilization | 1.413059 |
| 2 | CheatingBehavior | 1.110249 |
| 3 | Unwilingness_of_Responsibility | 1.259738 |
| 4 | StudentsHesitancyImpact | 1.332789 |
| 5 | TechDifficultyImpact | 0.979796 |
| 6 | HaveNetAccess | 1.273273 |
| 7 | HaveElectricSupply | 1.309307 |
| 8 | InteractionWihTeacher | 1.295518 |
| 9 | ClassParticipationChance | 1.286032 |
| 10 | AttensionAndFocusDisturb | 1.392692 |
| 11 | OnlineAndOfflineEqual | 1.214202 |
| 12 | TechnicalIssueImpact | 0.797957 |
| 13 | EconomicIssueImpact | 0.956183 |
| 3 | TeacherVoiceIssue | 1.027777 |
| 4 | LessTSInteraction | 1.045886 |

**TABLE II:**

| TTest Outcomes |
|---|
| Ttest_1sampResult(statistic=array([16.663333 , 31.38507589, 23.81939622, 19.65311057, 13.28871279, 11.51311097, 11.95187108, 13.35711613, 10.35574591, 11.7564528 , 13.11574349, 13.84994208, 13.79421828, 12.0044142 , 9.78640192, 10.375 , 13.09261879, 11.51416659, 13.36038922, 13.88838218, 13.86128572, 9.80912102, 18.24871239, 13.57080199, 12.19631092, 11.18462458, 10.35381536, 12.18694645, 13.15416906, 11.9272551 , 11.34226868, 10.64348064, 11.2720409 ]), pvalue=array([6.29551067e-18, 1.08636299e-26, 8.60469002e-23, 3.85366635e-20, 5.07753770e-15, 2.80770597e-13, 1.00580609e-13, 4.38220401e-15, 4.72979440e-12, 1.58421613e-13, 7.38588068e-15, 1.53985258e-15, 1.73086219e-15, 8.90861219e-14, 2.02190243e-11, 4.50637008e-12, 7.76732461e-15, 2.80069994e-13, 4.35148773e-15, 1.42079500e-15, 1.50369222e-15, 1.90647656e-11, 3.87615125e-19, 2.77545554e-15, 5.73530395e-14, 6.15085107e-13, 4.75281129e-12, 5.85928814e-14, 6.79392859e-15, 1.06477004e-13, 4.21455437e-13, 2.30654437e-12, 4.98568395e-13])) |

## VIII. CONCLUSTION

To evaluate and find the correctness and applicability of the hypothesis as per the problem statement, we used an online survey approach using Google docs. According to the percentages of survey responses given in Fig.1 Fig.2, Fig.3(a), Fig.3(b) and Fig. 4, availability of Laptop/Computer at student homes was 85.3% and smart phones was 87.5%. Time spent on social media during the online lecture was 71%. The level of Class participation was 49.6%. The concern of students cheating during the online exam was 70.8%—level of cheating behavior to ignore interest in online due to online exams encouraged by 60.8% of students. The student's unwillingness was found at 73.2%. Impacts of technical issues during online classes were 84.4% . The pace of the teacher's voice due to the Net problem was discovered at 79.5% and Impacts of less interaction of teacher-student found to be 78%. As per Fig. 4, psychological impacts on learning participation during online classes were discovered at 73.5% , the stress of loneliness affects students' level of involvement was 68.5% and the anxiety levels disturb students' level of motivation by 77.9%. As per the above experiments, the means and standard deviations are given in Table I and Table II above. Most of the high values of means shows that much percentage of the students are not fully involved during online lecture. Similarly, most of the values of standard deviations are far from zero. It shows that data points are far from the mean. We applied the test here because we don't have actual population means. We used ttest_1samp (Dataset [:35],0) with a default value of 0 to compute the variables' statistics and p-value. The results of this test are provided in Table II. These values are minimal, which is in favor of rejecting the null or H0 hypothesis and accepting the alternate or H1 hypothesis. It further means that students' involvement during online classes is severely affected.

## ACKNOWLEDGMENT

The authors are very grateful to the management of server room of Faculty of Computing and Information Technology (FCIT), University of the Punjab to forward our questionnaire to the students for the responses. We are also thankful to the Students of Undergraduates and Gradates students of FCIT for the warm participation and sincere responses during the survey of this research study.

## REFERENCES

[1]. Fernando, R., "The COVID-19 Pandemic: A call for a reality check.", published in Galle Medical Journal, 25 (1). 2020.

[2]. Myers, A., "After COVID-19: Recalibrating the American educational system", Retrieved from https://hub.jhu.edu/2020/04/07/bob-balfanz-education-reform-covid-19/, accessed 2021.

[3]. Tam, G., & El-Azar, D. (2020), "3 ways the coronavirus pandemic could reshape education", Retrieved from https://www.weforum.org/agenda/2021/05/consumer-demand-covid-19-recovery/ published in 2020 but accessed 2021.

[4]. World Bank, "How countries are using edtech (including online learning, radio, television, texting) to support access to remote learning during the COVID-19 pandemic.", Retrieved from https://www.worldbank.org/en/topic/edutech/brief/how-countries-are-using-edtech-to-support-remote-learning-during-the-covid-19-pandemic , published in 2020 but accessed 2021.

[5]. Ractham, P., & Chen, C., "Promoting the use of online social technology as a case-based learning tool", published in Journal of Information Systems Education, 24 (4), 2019.

[6]. Noor Ul Ain, “Is online education the new future of Pakistan?” Retrieved from https://dailytimes.com.pk/579663/blessing-in-disguise-is-online-education-the-new-future-of-pakistan/ , published in 2020 but accessed 2021.

[7]. Noor Ul Ain Ali., “Students disappointed with online teaching system amid COVID-19”. Retrieved from https://dailytimes.com.pk/587446/students-disappointed-with-online-teachingsystem-amid-covid-19/, published in 2020 but accessed 2021.

[8]. Kebritchi, M., Lipschuetz, A., & Santiague, L., “Issues and challenges for teaching successful online courses in higher education: A literature review”, Journal of Educational Technology Systems, 46 (1), 4–29, 2017.

[9]. Kunal Chaturvedi, Dinesh Kumar Vishwakarma, Nidhi Singh, “ COVID-19 and its impact on education, social life and mental health of students: A survey”, published in Children and Youth Services Review 121 (2021) 105866, pp. 1-6.

[10]. Nico W. Van Yperen, Melvyn R.W. Hamstra and Marloes van der Klauw, “ To Win, or Not to Lose, At Any Cost: The Impact of Achievement Goals on Cheating”, published in British Journal of Management, Vol. 22, S5–S15 (2011)

[11]. Lia M. Daniels, Lauren D, Goeganx Patti C. Parker, “ The impact of COVID 19 triggered changes to instruction and assessment on university students’ self reported motivation, engagement and perceptions”, published in Social Psychology of Education (2021) 24:299–318, Springer.

[12]. Timon ElmerID, Kieran Mepham, Christoph Stadtfeld, “Students under lockdown: Comparisons of students’ social networks and mental health before and during the COVID-19 crisis in Switzerland”, published in PLOS ONE, pp: 1-22, 2020.

[13]. Mohammed Amin Almaiah, Ahmad Al-Khasawneh and Ahmad Althunibat, “ Exploring the critical challenges and factors influencing the E-learning system usage during COVID-19 pandemic”, published in Journal of Education and Information Technologies, Springer.

[14]. Huma Sarwar, Hira Akhtar, Meshal Muhammad Naeem, Javeria Ali Khan, Khadija Waraich, Sumaiya Shabbir, Arshad Hasan and Zohaib Khurshid, “ Self-Reported Effectiveness of e-Learning Classes during COVID-19 Pandemic: A Nation-Wide Survey of Pakistani Undergraduate Dentistry Students”, published in European Journal of Dentistry, 2020;14(suppl S1):S34–S43.

[15]. Aqsa Arshad, Madiha Afzal, * Dr. Muhammad Sabboor Hussain, “ Sudden Switch to Post-COVID-19 Online Classes and Cognitive Transformation of ESL Learners: Critical Analysis of Discourse of Fear”, published in Research Journal of Social Sciences & Economics Review, Vol. 1, Issue 3, 2020, PP: 188-199.

[16]. Najmul Hasan, Yukun Bao, “ Impact of “e-Learning crack-up” perception on psychological distress among college students during COVID-19 pandemic: A mediating role of “fear of academic year loss”, published in Children and Youth Services Review 118 (2020) PP: 1-10.

[17]. Muhammad Anwar, Anwar Khan, Khalid Sultan, “The barriers and challenges faced by students in online education during covid-19 pandemic in Pakistan”, published in Gomal University Journal of Research, Volume 36, Issue 1, JUNE, 2020. PP. 52-62.

[18]. Brochure, what is a survey? Bill Kalsbeek, 1995 publications officer, ASA section on survey research methods, 1995.